\documentclass[10pt, letterpaper]{article}
\usepackage[OME]{express}
\usepackage{color}
\usepackage{soul}%This package will be deleted before submission
\usepackage[normalem]{ulem}
\usepackage{cancel}

\usepackage{latexsym, amssymb, amsfonts, euscript, amsmath}
\usepackage{multirow}

\newcommand{\abs}[1]{\left\vert#1\right\vert}

\newcommand{\pdet}{p_{\!_{_\mathrm{D}}}}
\newcommand{\pfr}{p_{_\mathrm{fr}}}

\newcommand{\phiSample}{\Phi_{\mathrm{s}}}
\newcommand{\p}[1]{p_{_#1}}

\begin{document}
%\title{Simple template for authors submitting to OSA Express Journals}

%%%%%%%%%%%%%%%%%% title page information %%%%%%%%%%%%%%%%%%

\title{Sample phase gradient and fringe phase shift in dual phase grating X-ray interferometry }

\author{Aimin Yan,\authormark{1} Xizeng Wu,\authormark{1,*} and Hong Liu\authormark{2}}

\address{\authormark{1}Department of Radiology, University of Alabama at Birmingham, Birmingham, AL 35249, USA \\
\authormark{2}Center for Bioengineering and School of Electrical and Computer Engineering, University of Oklahoma, Norman, OK 73019, USA}

\email{\authormark{*}xwu@uabmc.edu} %% email address is required

% \homepage{http:. . . } %% author's URL, if desired

%%%%%%%%%%%%%%%%%%% abstract and OCIS codes %%%%%%%%%%%%%%%%
%% [use \begin{abstract*}. . . \end{abstract*} if exempt from copyright]

\begin{abstract}
One of the key tasks in grating based x-ray phase contrast imaging is to accurately retrieve local phase gradients of a sample from measured intensity fringe shifts. To fulfill this task in dual phase grating interferometry, one needs to know the exact mathematical relationship between the two. In this work, using intuitive analysis of the sample-generated fringe shifts  based on the beat pattern formation mechanism, the authors derived the formulas relating sample's phase gradients to fringe phase shifts.  These formulas provide also a design optimization tool for dual phase grating interferometry. 
\end{abstract}


%%%%%%%%%%%%%%%%%%%%%%% References %%%%%%%%%%%%%%%%%%%%%%%%%
%\begin{thebibliography}{99}[ref?]

%\bibitem{gallo99} K. Gallo and G. Assanto, ``All-optical diode based on second-harmonic generation in an asymmetric waveguide, '' \josab {\bfseries 16}(2), 267--269 (1999). 

%\end{thebibliography}

%\bibliographystyle{osajnl}
%\bibliography{ref}

\section{Introduction}\label{sec-intr}
In x-ray interferometry based phase contrast imaging, one of  the key tasks is to  retrieve local phase gradients of a sample. More specifically, the sample's local refraction angle, $\alpha(x, y)$, is proportional to the local phase gradient of the sample\cite{Momose, Weitkamp, Momose-2, Pfeiffer}:
\begin{equation}\label{eq-alpha}
\alpha(x, y) = \frac{\lambda}{2\pi}\cdot\frac{\partial\phiSample(x, y)}{\partial x}, 
\end{equation}
where $\lambda$ is x-ray wavelength and $\phiSample(x, y)$ denotes local sample phase shift, which is equal to  $\phiSample(x, y)=-\lambda r_{\mathrm{e}}\int \rho_{\mathrm{e}}(x, y, s)ds$. In this integral $\rho_{\mathrm{e}}$ denotes the sample's electron density and $r_{\mathrm{e}}$ is the classical electron density which equals to $2.82\times 10^{-15}$m. In Eq.~(\ref{eq-alpha}), $\partial\phiSample(x, y)/\partial x$ denotes sample's phase gradient along the direction perpendicular to grating slits.  In x-ray interferometry one measures the fringe shift $\Delta\phi$ generated by sample refraction. In order to retrieve local phase gradients of the sample, one must find out functional relationship between intensity fringe shifts and sample phase gradients. It turns out that the mathematical relation between them depends not only on the interferometer's geometrical configuration and phase grating periods, but also on the position of the sample.  So one important task in x-ray interferometry is to rigorously derive this functional relationship  which will enable retrieval of sample phase gradients from measured interference fringe shifts. Combined with tomography, the retrieved sample gradients from angular projections can be used to reconstruct 3D maps of sample electron densities. 

\begin{figure}[htbp]
\centering\includegraphics[width=\textwidth]{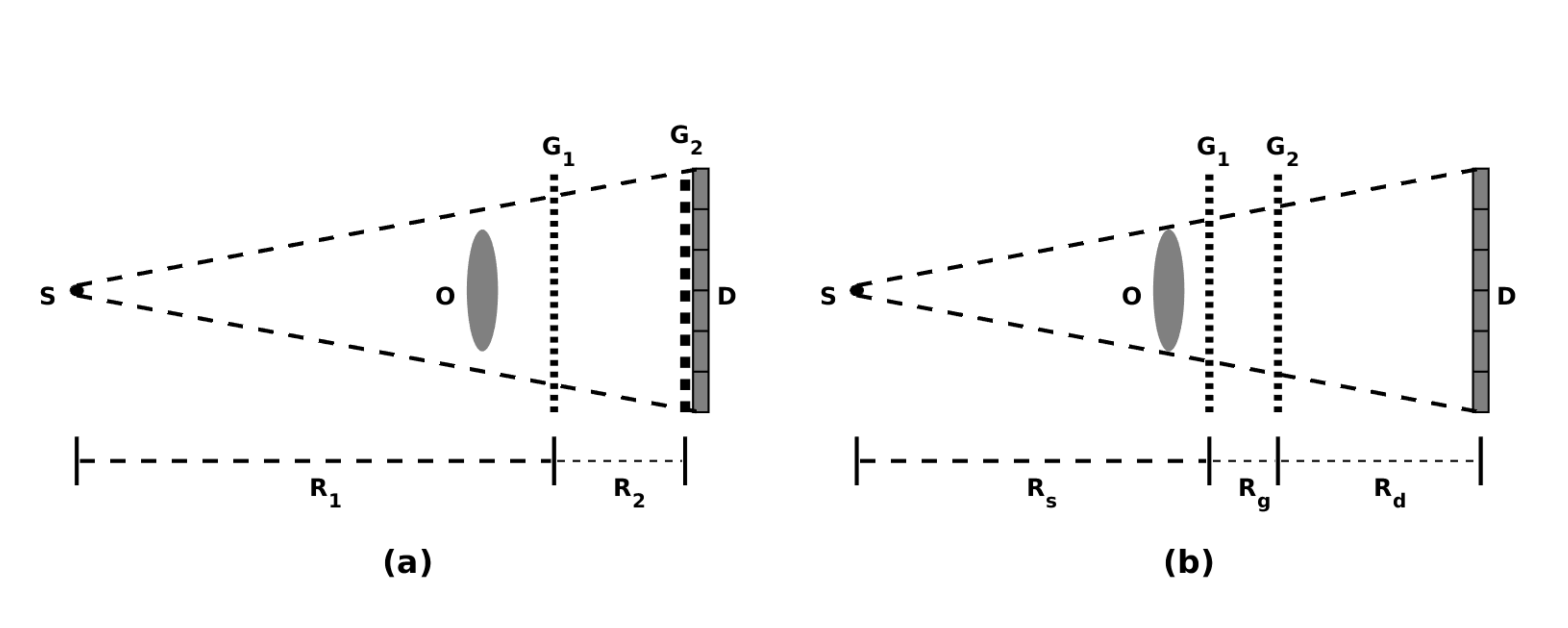}
\caption{Schematic of Talbot-Lau interferometry (a), and dual-phase grating interferometry (b). } \label{fig-setup}
\end{figure}
In Talbot-Lau interferometry, where only a single phase grating is employed as the beam splitter, thereby high-contrast interference fringes form at certain distances downstream (Fig.~\ref{fig-setup}(a)). Sample refraction distorts the intensity fringe pattern, and the mathematical relations between fringe shifts and sample phase gradient is well described in literature, and the specific  formulas depend on whether the sample is placed upstream or downstream of the phase grating\cite{Momose, Weitkamp, Momose-2, Pfeiffer}. 

\begin{equation}\label{eq-phi-1g}
\Delta\phi(x, y)=\left\{\begin{array}{ll}
\frac{\lambda R_1}{(R_1+R_2)p_{_\mathrm{eff}}} L_{\mathrm{D}}\cdot\frac{\partial\phiSample(x, y)}{\partial x}, &\mathrm{if\;Sample\;at\;downstream\;of\;} G_1,\\
\frac{\lambda R_2}{(R_1+R_2)p_{_\mathrm{eff}}} L_{\mathrm{S}}\cdot\frac{\partial\phiSample(x, y)}{\partial x}, &\mathrm{if\;Sample\;at\;upstream\;of\;} G_1,
\end{array}\right.
\end{equation}
where $R_1$  is the source-to-phase grating ($G_1$) distance, and $R_2$  the distance from $G_1$ to detector entrance as is shown in Fig.~\ref{fig-setup}(a). Note that the absorbing grating $G_2$ in Fig.~\ref{fig-setup}(a) serves as an analyzer for resolving the intensity fringes. In Eq.~(\ref{eq-phi-1g}),  $p_{_\mathrm{eff}}$ denotes the effective grating period of the phase grating $G_1$,  $p_{_\mathrm{eff}}=\p1/2$ if it is a $\pi$-grating, and $p_{_\mathrm{eff}}=\p1$ otherwise. In Eq.~(\ref{eq-phi-1g}) $L_{\mathrm{D}}$ denotes sample-to-detector distance, and $L_{\mathrm{S}}$ the sample-to-source distance. However, to measure fringe phase shifts with common image detectors, one usually has to  use a fine-pitch absorbing grating placed at the detector's entrance. One indirectly detects fringe pattern through grating scanning, which is also called phase stepping procedure\cite{Momose, Weitkamp, Momose-2, Pfeiffer, Yashiro, Zhu-Zhang, Bevins, Tang-Yang-2, Bennett, Suortti, Momose-Huwakara, Morimoto-Fujino, Morimoto, Y-W-L, Y-W-L-2}. The absorbing grating blocks more than half of transmitting x-ray, and will significantly increase radiation dose in imaging exams.  This is a serious disadvantage of Talbot-Lau interferometry for radiation-dose sensitive imaging applications such as medical imaging.  

Recently, dual phase grating x-ray interferometry emerges as an attractive alternative\cite{Miao, Kagias, Y-W-L-dual-grating-1, Bopp}. A typical setup of dual phase grating interferometers employs two phase gratings $G_1$ and $G_2$ as the beam splitters, as is shown in Fig.~\ref{fig-setup}(b). The split waves transmitting through the phase gratings  create different diffraction orders interfering with each other.  The intensity fringe pattern includes a beat pattern\cite{Y-W-L-dual-grating-1}. The imaging detector $D$ has a pixel size much larger than periods of both the phase gratings (Fig.~\ref{fig-setup}(b)). The detector just resolves the beat patterns of large periodicities, and renders other fine patterns to a constant background. Hence, different from Talbot-Lau interferometry\cite{Momose, Weitkamp, Momose-2, Pfeiffer, Yashiro, Zhu-Zhang, Bevins, Tang-Yang-2, Bennett, Suortti, Momose-Huwakara, Morimoto-Fujino, Morimoto, Y-W-L, Y-W-L-2}, the dual phase grating interferometers directly detect interference fringes without the need of absorbing analyzer grating. This advantage brings significant radiation dose reduction as compared to Talbot-Lau interferometry.

However, to retrieve local phase gradients of a sample, those formulas  in Eq.~(\ref{eq-phi-1g})  are not applicable to dual phase grating interferometers, in which refracted x-rays may pass through two phase-gratings rather than a single phase grating.  In this work we set out to derive the corresponding formulas for dual phase grating x-ray interferometry.  In section~\ref{sec-method}, using intuitive analysis of the sample-generated fringe shifts and the intensity beat pattern formation,  we derived  formulas that relate the sample's phase gradients to fringe phase shifts.   In section~\ref{sec-results} we perform wave propagation simulations to validate the formulas derived in section~\ref{sec-method}. We also show how the angular sensitivity  of a dual phase grating interferometer is determined in dual phase grating interferometry. We conclude the paper in section~\ref{sec-conclude}.

\section{Methods}\label{sec-method}
We start our derivation from explaining the fringe formation mechanism in dual phase grating interferometry. Figure~\ref{fig-setup}(b) shows the geometrical configuration of a dual phase grating interferometer.  It consists of a source $S$, two phase gratings $G_1$ and $G_2$, and an imaging detector $D$. The periods of the first and second phase gratings are $\p1$ and $\p2$ respectively, $R_s$ is the source-to-$G_1$ distance. Note that $R_g$ is the spacing between the two phase gratings, and $R_d$ denotes the $G_2$-to-detector distance (Fig.~\ref{fig-setup}(b)).  For sake of convenience in discussion, we define several magnification factors as follows:
\begin{equation}\label{eq-Ms}
M_{g_1}=\frac{R_s+R_g+R_d}{R_s}; \qquad M_{g_2}=\frac{R_s+R_g+R_d}{R_s+R_g},
\end{equation}
where $M_{g_1}$ represents the geometric magnification factor from $G_1$ to detector plane, and $M_{g_2}$ is the geometric magnification factor from $G_2$ to detector plane. Obviously, in absence of $G_2$-grating, a diffraction order of the intensity pattern generated by $G_1$ phase grating alone would be represented by $\exp\left[i2\pi (l\cdot x)/(M_{g_1}\cdot\p1)\right]$, where $l$ is an integer and indexes the diffraction order.  Similarly, in absence of $G_1$-grating, a diffraction order of the intensity pattern generated by $G_2$ grating alone would be represented by $\exp\left[i2\pi(m\cdot x)/(M_{g_2}\p2)\right]$, where $m$ indexes the diffraction order of the fringe in absence of $G_1$-grating. As is shown from our theory of dual phase grating interferometry\cite{Y-W-L-dual-grating-1}, x-ray irradiance at detector entrance is a result of cross-modulation between the fringe patterns generated by phase gratings $G_1$ and $G_2$ respectively. Hence the irradiance pattern at the detector entrance is a weighted sum of different diffracted orders. Each of the orders in dual phase grating interferometry is indexed by two integers $(l, m)$ and represented by a product as:
\begin{equation}\label{eq-cross-prod}
\exp\left[i2\pi \frac{l\cdot x}{M_{g_1}\p1}\right] \cdot \exp\left[i2\pi \frac{m\cdot x}{M_{g_2}\p2}\right]=\exp\left[i2\pi  x\cdot\left(\frac{l}{M_{g_1}\p1}+\frac{m}{M_{g_2}\p2}\right)\right].
\end{equation}
However, among these diffracted orders, there are beat patterns formed by those diffraction orders characterized by $l=-m$. With proper setup such that $1/(M_{g_1}\p1)$ close to $1/(M_{g_2}\p2)$ (for example $R_g\ll R_s, R_d$ and $\p1\approx \p2$), the period of these beat patterns can be made much larger than $\p1$ and $\p2$. As long as the detector pitch $\pdet\gg\p1, \p2$,  the detector-resolved intensity pattern is just the beat pattern, since the detector renders all other fine fringes of those $l\neq -m$ orders to a constant background. The beat pattern consists of different harmonics resolved by the imaging detector. The period of the fundamental mode of the beat patterns is\cite{Y-W-L-dual-grating-1, Miao}: 
\begin{equation}\label{eq-pfr}
\pfr = \frac{1}{1/(M_{g_2}\p2)-1/(M_{g_1}\p1)} = \frac{R_s+R_g+R_d}{R_g/\p2+R_s(1/\p2-1/\p1)}.
\end{equation}
Obviously, the period of the $m$-th harmonics is  $\pfr/m$. According to Eq.~(\ref{eq-pfr}), the period of the beat pattern can be much larger than grating periods $\p1$ and $\p2$, provided that the inter-grating spacing $R_g$ is set much smaller than the distances $R_s$ and $R_g$. With large fringe periods the beat pattern may be resolved by a common imaging detector of pixels in few tens of micrometers.

\begin{figure}[htbp]
\centering\includegraphics[width=0.9\textwidth]{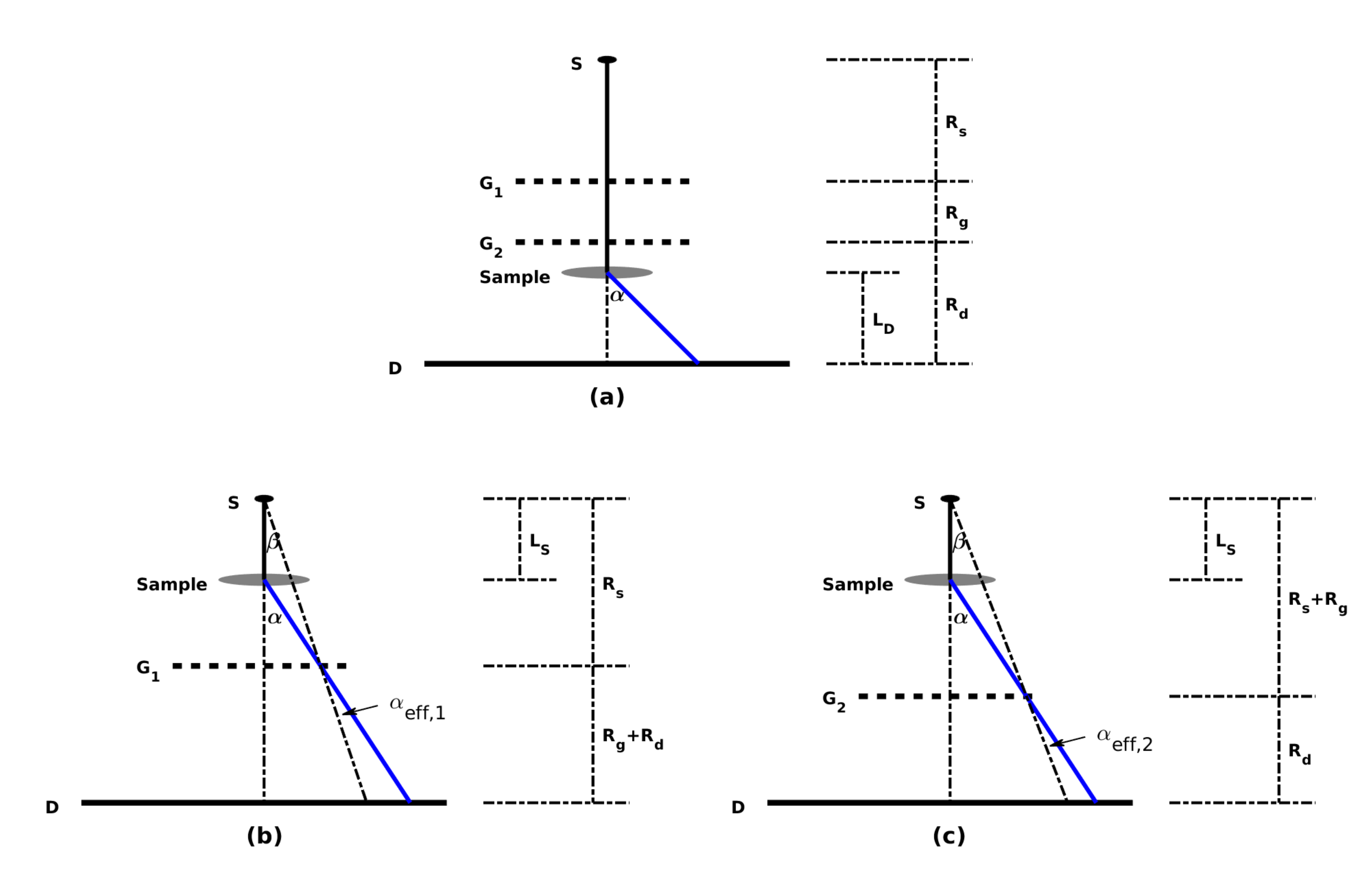}
\caption{Geometric configurations for fringe shift analysis. (a) Sample positioned downstream of $G_2$. (b) and (c) Sample positioned upstream of $G_1$. In (b) $G_2$ is assumed absent. In (c) $G_1$ is assumed absent. } \label{fig-struct}
\end{figure}

Once the fringe formation mechanism is understood, we are now ready to derive the mathematical relationship between fringe phase shift and sample phase gradient. First, we consider a relatively simple case of sample being downstream of the second phase grating $G_2$. Let a sample's local refraction angle be $\alpha(x, y)$,  and the sample-detector distance be $L_{\mathrm{D}}$. Referring to Fig.~\ref{fig-struct}(a), the fringe lateral shift caused by the refraction is equal to $\alpha L_{\mathrm{D}}$. Since the period of the $m$-th diffracted order in the beat pattern is $\pfr/m$, using Eq.~(\ref{eq-alpha}) and Eq.~(\ref{eq-pfr}) we found the corresponding fringe phase shift $\Delta\phi_{\mathrm{down}}(x, y, m)$ as
\begin{equation}\label{eq-phi-down}
\Delta\phi_{\mathrm{down}}(x, y;m)=\frac{2\pi\alpha(x, y)L_{\mathrm{D}}}{\pfr/m}=m\lambda \frac{L_{\mathrm{D}}}{\pfr}\cdot\frac{\partial \phiSample(x, y)}{\partial x}.
\end{equation}
In above equation the subscript in $\Delta\phi_{\mathrm{down}}(x, y;m)$ indicates that the sample is positioned downstream of the second grating. Equation~(\ref{eq-phi-down}) shows that the fringe phase shift scales with sample phase gradient,  and the proportional constant is given by $m\lambda L_{\mathrm{D}}/\pfr$. In practice, the dominant diffracted order for $\pi/2$-gratings is $m=1$ and that for $\pi$-gratings is $m=2$.

On the other hand, there is a different way to place the sample in the beam. In dual phase grating interferometry, one does not place sample between two phase gratings, since the spacing between them can be as narrow as few millimeters. Therefore, another way of sample placement is to place the sample upstream of the first phase grating $G_1$. Assume that the sample is placed upstream of $G_1$ and the source-sample distance is $L_{\mathrm{S}}$. Let us firstly consider how the sample phase gradient would affect $G_1$-associated fringe in absence of $G_2$ grating. Consider a ray refracted by the sample with an angle $\alpha$. Referring to Fig.~\ref{fig-struct}(b), the refracted ray propagates to $G_1$ grating, and the refracted ray makes an angle $\alpha_{\mathrm{eff}, 1}$ to the locally undisturbed ray. We call $\alpha_{\mathrm{eff}, 1}$ the effective refraction angle. In practice,  sample refraction angles are much smaller than 1 radian. Based on the geometry depicted in Fig.~\ref{fig-struct}(b), under assumption of $\alpha \ll 1$, it is easy to find that $\alpha_{\mathrm{eff}, 1}=(L_{\mathrm{S}}/R_s)\alpha$. That is, compared to the sample refraction angle, the effective refraction angle is reduced by a factor, which is the ratio of source-to sample distance over the sample to grating distance. Propagating over a distance $R_g+R_d$ to the detector, in absence of $G_2$ grating, this refracted ray would cause a lateral shift of the $G_1$-associated fringe by $\alpha_{\mathrm{eff}, 1}\cdot(R_g+R_d) = (L_{\mathrm{S}}/R_s)\cdot (R_g+R_d) \alpha$. It is easy to see from Eq.~(\ref{eq-cross-prod}) that in absence of $G_2$ grating the period of the $l$-th order $G_1$-fringe is $M_{g_1}\p1/l$, hence the $G_1$-fringe phase shift caused by the sample refraction is equal to:
\begin{equation}\label{eq-phi-g1}
\Delta\phi_{G_1}(l) = 2\pi l  \frac{L_{\mathrm{S}}}{R_s}\cdot \frac{R_g+R_d}{M_{g_1}\p1}\cdot\alpha.
\end{equation}

A similar derivation can be applied to $G_2$-fringe phase shift, assuming absence of $G_1$  grating. Referring to Fig.~\ref{fig-struct}(c), the effective refraction angle is $\alpha_{\mathrm{eff}, 2}$. As compared to the refraction angle $\alpha$, it should be reduced by the ratio of source-to-sample distance over the sample to grating distance, thereby $\alpha_{\mathrm{eff}, 2}$ becomes $\alpha_{\mathrm{eff}, 2}=L_{\mathrm{S}}/(R_s+R_g)\alpha$. Propagating over a distance $R_d$ to the detector, this refracted ray would cause a lateral shift of the $G_2$-associated fringe by $\alpha_{\mathrm{eff}, 2}\cdot R_d=(L_{\mathrm{S}}/(R_s+R_g))\cdot R_d\alpha$. Equation~(\ref{eq-cross-prod}) shows that in absence of $G_1$ grating the period of the $m$-th order $G_2$-fringe is $M_{g_2}\p2/m$. Hence, the sample refraction generates a phase shift for the $m$-th order $G_2$-fringe:
\begin{equation}\label{eq-phi-g2}
\Delta\phi_{G_2}(m) = 2\pi m \frac{L_{\mathrm{S}}}{R_s+R_g}\cdot\frac{R_d}{M_{g_2}\p2}\cdot\alpha.
\end{equation}
In dual phase grating interferometry the interference fringe pattern interferometry is indexed by two integers $(l, m)$ and represented by a product as $\exp\left[i2\pi (l\cdot x)/(M_{g_1}\cdot\p1)\right]\times \exp\left[i2\pi (m\cdot x)/(M_{g_2}\cdot\p2)\right]$,   thus  the refraction-generated  fringe  phase shift of  the order $(l, m)$ is 
\begin{equation}\label{eq-phi-2g}
\Delta\phi_{\mathrm{up}}(l, m) = \Delta\phi_{G_1}(l)+\Delta\phi_{G_2}(m).
\end{equation}

In dual phase grating x-ray interferometry, one usually adopts a common imaging detector of pixels much larger than grating periods.  As is discussed in section~\ref{sec-method} on fringe formation mechanism, due to the pixel averaging effects, the detector resolves only the beat patterns of diffracted orders $(l=-m, m)$. This being so, Eq.~(\ref{eq-phi-2g}) gives the sample-refraction generated fringe phase shift for the $m$-th order beat patterns as follows:  
\begin{equation} \label{eq-phi-up-middle}
\Delta\phi_{\mathrm{up}}(m) = \Delta\phi_{G_1}(-m)+\Delta\phi_{G_2}(m)=-2\pi m \cdot \frac{L_{\mathrm{S}}}{\p0}\alpha,
\end{equation}
where
\begin{equation}\label{eq-p0}
\p0=\frac{R_s+R_g+R_d}{R_g/\p1+R_d (1/\p1-1/\p2)}.
\end{equation}
Comparing Eqs.~(\ref{eq-pfr}) and (\ref{eq-p0}), one can see $\p0$ is always equal to $\pfr$ if the two gratings have the same period ($\p1=\p2$). But for $\p1\neq\p2$, $\abs{\p0}$ can be greater or less than $\abs{\pfr}$, depending on the geometric setup. More explicitly, we rewrite the sample-refraction generated fringe phase shift for the $m$-th order beat patterns as:
\begin{equation}\label{eq-phi-up}
\Delta\phi_{\mathrm{up}}(x, y;m)=-m\lambda \frac{L_{\mathrm{S}}}{\p0}\cdot\frac{\partial \phiSample(x, y)}{\partial x}.
\end{equation}
In above equation the subscript in $\Delta\phi_{\mathrm{up}}(x, y;m)$ indicates that the sample is positioned upstream of the first grating $G_1$. Equation~(\ref{eq-phi-up}) shows that, when a sample is positioned upstream of the $1^{\mathrm{st}}$ phase grating,  the fringe phase shift scales again with sample phase gradient, and the proportional constant is $-m \lambda L_\mathrm{S}/\p0$. Especially, the closer to the $1^{\mathrm{st}}$ phase grating the sample is, the larger magnitude the fringe phase shift is. 

Obviously, this proportional constant in Eq.~(\ref{eq-phi-up}) is different to that in Eq.~(\ref{eq-phi-down}), which corresponds to the setups with sample positioned downstream of the $2^{\mathrm{nd}}$ phase grating.

\section{Results}\label{sec-results}

\begin{figure}[h]
\begin{center}
				\includegraphics[width=0.9\textwidth]{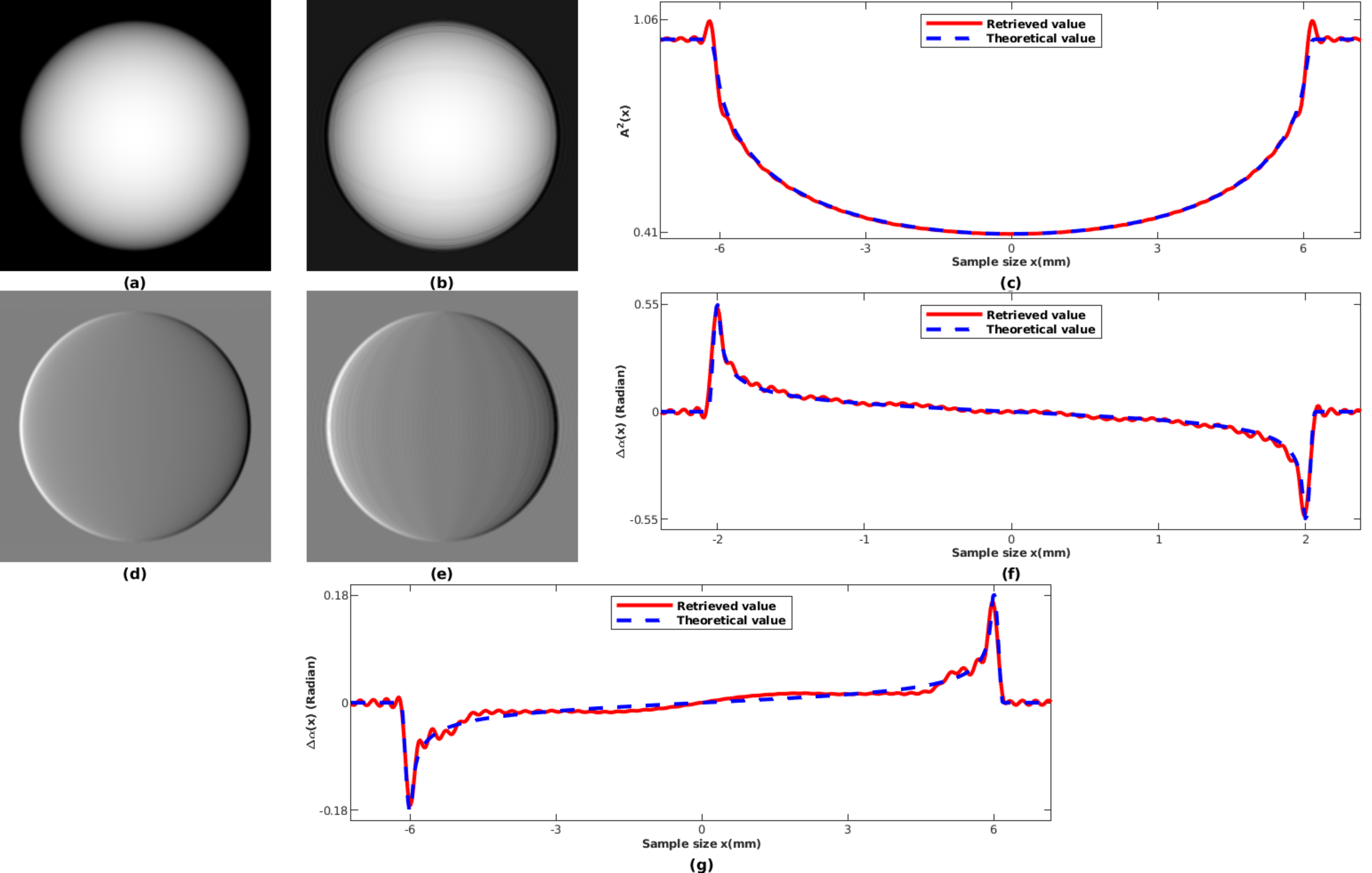}
    \caption{Simulation results for validation of Eq.~(\ref{eq-phi-dual-g}). In the simulation, the setup consists of two $\pi$-gratings and a 20keV point source.  A sphere shaped sample of adipose tissue is placed upstream of the $1^{\mathrm{st}}$ grating and downstream of the $2^{\mathrm{nd}}$ grating respectively.  We employed ray-tracing  to compute the theoretical map of  sample's attenuation $A^2(x, y)$ and fringe phase shift $\Delta\phi(x, y)$. We then simulated Fresnel wave propagation in the interferometer   to generate  the intensity fringe patterns. The fringe phase shift $\Delta\phi(x, y)$ and attenuation map $A^2(x, y)$ are retrieved through the Fourier method.  The ray-tracing generated theoretical maps of attenuation $A^2(x, y)$ and fringe phase shift $\Delta\phi(x, y)$, when sample is place half way between the source and the $1^{\mathrm{st}}$ grating, are shown in Fig.~3(a) and (d); while the retrieved $A^2(x, y)$ and $\Delta\phi(x, y)$ are shown in Fig.~\ref{fig-p1=p2}(b) and (e) respectively.  For the purpose of comparison, the profiles  of the attenuation and fringe phase shift along the central line across grating are shown  in Fig.~\ref{fig-p1=p2}(c) and (f). The blue curves correspond to the theoretical values, while the red curves correspond to the values retrieved from intensity fringes.  Figure~\ref{fig-p1=p2}(g) is the profiles of the fringe phase shift $\Delta\phi(x)$ along the central line across grating, when the sample is placed to the plane half way between $G_2$ grating and detector. The good agreement between the theoretical values and the fringe-pattern retrieved values provide a validation of Eq.~(\ref{eq-phi-dual-g}). For details, see text.} 
    \label{fig-p1=p2}
\end{center}
\end{figure}

Combining Eqs.~(\ref{eq-phi-down}) and (\ref{eq-phi-up}) together,  we found the measured fringe phase shift $\Delta\phi$ is proportional to sample's phase gradient with a proportional constant $\xi$ given by
\begin{equation}\label{eq-phi-dual-g}
\left.\begin{array}{ll}
\displaystyle{\Delta\phi(x, y) = \xi \cdot\frac{\partial \phiSample(x, y)}{\partial x},}\vspace{2pt}\\
\displaystyle{\xi=}\left\{
\begin{array}{rl}
\displaystyle{-m\lambda\frac{L_\mathrm{S}}{\p0},} &\displaystyle{\mathrm{if\; Sample\;is\; upstream\;of\;}G_1,}\vspace{4pt}\\
\displaystyle{m\lambda\frac{L_\mathrm{D}}{\pfr},} &\displaystyle{\mathrm{if\; Sample\;is\; downstream\;of\;}G_2,}
\end{array}\right.
\end{array}\right.
\end{equation}
where $m=2$ if the phase gratings are $\pi$ gratings and $m=1$ otherwise. In Eq.~(\ref{eq-phi-dual-g}), $\p0$ and $\pfr$ are given in Eqs.~(\ref{eq-p0}) and (\ref{eq-pfr}) respectively.  Equation~(\ref{eq-phi-dual-g}) shows that this formula lays foundation of quantitative imaging for sample phase gradients with dual phase grating x-ray interferometry.  Based on our understanding of fringe formation mechanism\cite{Y-W-L-dual-grating-1}, we developed above intuitive and heuristic method to derive Eqs.~(\ref{eq-phi-dual-g}).

To validate the fringe shift formula of Eq.~(\ref{eq-phi-dual-g}), we compare the calculated fringe phase shift of formula Eq.~(\ref{eq-phi-dual-g}) to that determined through numerical simulations. We employed four different interferometer setups in the simulation study.  In the first simulation, a point x-ray source of $20$keV design energy and dual-$\pi$ phase gratings of period $\p1=\p2 = 1\mu$m are employed. The geometric setup is $R_s=R_d=450$mm and $R_g=5$mm. With this geometric setup, the fringe can attain a high visibility pattern and large fringe period $\pfr=\p0=181\mu$m as well\cite{Y-W-L-dual-grating-1}. So with a detector of $\pdet=36.2\mu$m pitch, one period of the resolved fringe will take up five pixels. To validate the fringe shift formula of Eq.~(\ref{eq-phi-dual-g}), a sample is placed half way between the source and $G_1$-plane, i.e. $L_{\mathrm{S}}=R_s/2=225$mm. The sample is assumed a sphere of diameter 4 mm filled with $100\%$ adipose tissue. We employed ray-tracing to get the projected sample attenuation $A^2(x, y)$,  and the phase map of the sample. From sample phase map we computed sample phase gradients. Using Eq.~(\ref{eq-phi-dual-g}) and the sample phase gradients we theoretically predict a map of the fringe shift $\Delta\phi(x, y)$. The theoretical map of sample attenuation $A^2(x, y)$ and map of fringe shift $\Delta\phi(x, y)$  are shown in Fig.~\ref{fig-p1=p2}(a) and (d) respectively. We then employed Fresnel propagation down stream from sample plane through the gratings $G_1$, $G_2$ and finally to the detector plane to get the projected image. We then employed the Fourier method for retrieval of the fringe phase shift $\Delta\phi(x, y)$ and attenuation map $A^2(x, y)$. The maps of retrieved attenuation and phase shift are shown in Fig.~\ref{fig-p1=p2}(b) and (e). As comparison, profiles of the attenuation $A^2(x)$ and phase shift $\Delta\phi(x)$ along the central line across grating are shown in Fig.~\ref{fig-p1=p2}(c) and (f) respectively. The dashed blue curves represent the theoretical value, while the solid red curves are the retrieved values from the intensity fringes. The good match between the theoretical and retrieved values of $\Delta\phi(x)$ validates the fringe phase shift formula of Eq.~(\ref{eq-phi-dual-g}) for sample upstream $G_1$. 

In the second simulation, we keep the same setup as in the previous simulation but placed sample to the half way between the $G_2$-grating and detector plane, i.e, $L_{\mathrm{D}}=R_d/2=225$mm. The diameter of the sample sphere is also increased to $12$mm to increase the resolution. The profiles of the phase fringe shifts $\Delta\phi(x)$ along the central line across grating are shown in Fig.~\ref{fig-p1=p2}(g), in which the dashed blue line represents the theoretical value and the solid red line is the retrieved one. The closeness of the two profiles  validates the fringe phase shift formula of Eq.~(\ref{eq-phi-dual-g}) for sample downstream $G_2$. Comparing Fig.~\ref{fig-p1=p2}(f) and (g) in previous page one finds the profiles of $\Delta\phi(x)$ are reflected, in other words, the values of $\Delta\phi(x)$ change signs when the sample moves from upstream to downstream, as is predicted by the sign difference in Eq.~(\ref{eq-phi-dual-g}) for the two scenarios.  Noting this interesting feature is important in practice for accurately retrieving sample phase gradients from measured fringe phase shifts.
\begin{figure}[hbtp]
\begin{center}
				\includegraphics[width=0.6\textwidth]{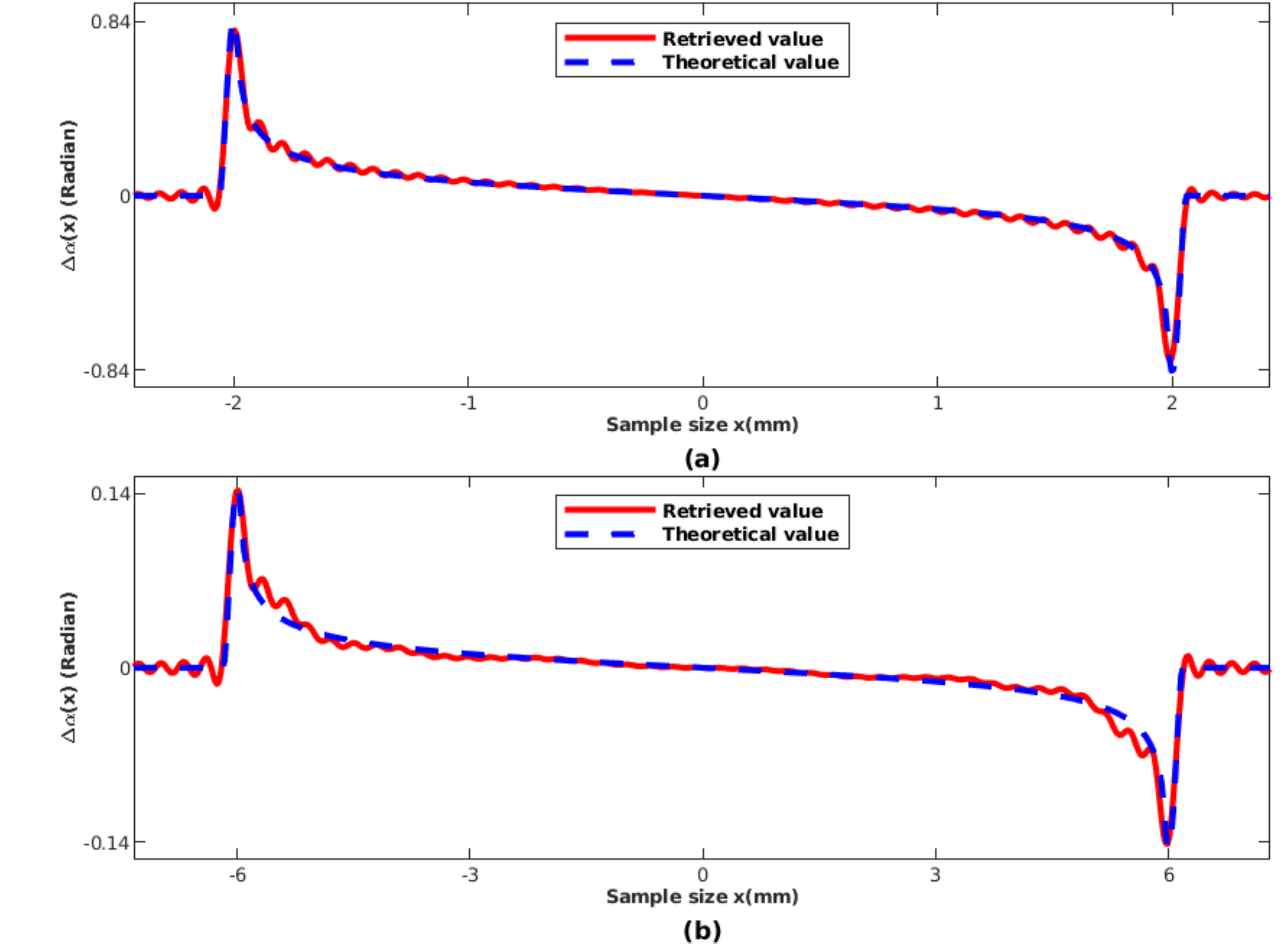}
    \caption{In this simulation, the second grating is replaced with a $\p2=1.1\mu$m period $\pi$ grating but the first grating is kept the same as that in Fig.~\ref{fig-p1=p2}. The geometric setup is also changed to $R_s=450$mm, $R_g=40$mm, and $R_d=380$mm and detector pixel pitch $\pdet=39.1\mu$m. By placing the sample to plane $L_{\mathrm{S}}=R_s/2=225$mm and to plane $L_{\mathrm{D}}=R_d/2=190$mm, one gets the retrieved plots of $\Delta\phi(x)$ for sample upstream of $G_1$ (Fig.~\ref{fig-p1_not=p2}(a)) and for sample downstream of $G_2$ (Fig.~\ref{fig-p1_not=p2}(b)).}
    \label{fig-p1_not=p2}
\end{center}
\end{figure}

In the third and fourth simulations, we replace the second phase grating $G_2$ with a $\p2=1.1\mu$m period  $\pi$-phase grating and keep the first grating $G_1$ the same. Corresponding to this change, the geometric setup is changed accordingly for attaining good fringe visibility\cite{Y-W-L-dual-grating-1}. So in the third and fourth simulations we set the interferometer with $R_s=450$mm, $R_d=380$mm, and $R_g=40$mm.  In this configuration,  the fringe period, as is determined by Eq.~(\ref{eq-pfr}), is $\pfr=-191.4\mu$m and $\p0=11.67\mu$m\cite{Y-W-L-dual-grating-1}. In addition,  the detector pixel size is changed  to  $\pdet=31.9\mu$m, thereby one fringe period can take up 6 pixels. We then simulate the fringe formation processes by placing the sample upstream to plane $L_{\mathrm{S}}=R_s/2=225$mm in the third simulation, and downstream to plane $L_{\mathrm{D}}=R_d/2=190$mm in the fourth simulation. The results are shown in Fig.~\ref{fig-p1_not=p2}. The profiles of $\Delta\phi(x)$ along the central line for sample upstream of $G_1$ are shown in Fig.~\ref{fig-p1_not=p2}(a) and those for sample downstream of $G_2$ are shown in Fig.~\ref{fig-p1_not=p2}(b), in which the dashed blue lines are the theoretical values and the solid red lines correspond to the retrieved ones. The close match between the red and blue lines  again validated Eq.~(\ref{eq-phi-dual-g}). 

In literature the ratio $\Delta\phi/(2\pi\alpha)$ is called the angular sensitivity.  It is a measure of fringe phase shift generated per unit sample-refraction angle. Apparently, the larger the $\Delta\phi/(2\pi\alpha)$ is, the larger the magnitude of the fringe phase shift is. Equation~(\ref{eq-phi-dual-g}) can be used to compute the angular sensitivity of a given setup for design optimization of an interferometer. For example, Tab.~\ref{tab-1} lists the angular sensitivity values computed by using Eq.~(\ref{eq-phi-dual-g}) for the two setups investigated in the $3^{\mathrm{rd}}$ and $4^{\mathrm{th}}$ simulations. The notations used for describing the setups are explained in the text.

This table shows clearly that the angular sensitivity of the setup with the sample placed upstream of the $1^{\mathrm{st}}$ phase grating is about nineteen times that of the setup with sample positioned downstream of the $2^{\mathrm{nd}}$ phase grating.  With a given geometric setup, the closer the sample is placed to the first grating for sample upstream or the closer the sample is to the second grating for sample downstream, the larger the angular sensitivity is. For given sample, the setup with a higher angular sensitivity will generate larger magnitude of fringe phase shifts for measurement.

\begin{table}[hbp]
\begin{center}
\caption{Angular sensitivities of the setups in simulations 3 and 4.}\label{tab-1}
\begin{tabular}{|c|c|c|}
\hline
&Simulation 3: & Simulation 4:\\
&(Sample upstream) &(Sample downstream)\\
\hline
$\p1 (\mu \mathrm{m})$ & 1 &1\\
\hline
$\p2 (\mu \mathrm{m})$ & 1.1 &1.1\\
\hline
$R_s (\mathrm{mm})$ & 450 &450\\
\hline
$R_g (\mathrm{mm})$ & 40 &40\\
\hline
$R_d (\mathrm{mm})$ & 380 &380\\
\hline
$L_s (\mathrm{mm})$ & 225 &--\\
\hline
$L_d (\mathrm{mm})$ & -- &190\\
\hline
Angular sensitivity & 3.86E+04 &1.99E+03\\
\hline
\end{tabular}
\end{center}
\end{table}

Carefully examining Fig.~\ref{fig-p1_not=p2}(a) and (b), one may notice that the values of $\Delta\phi(x)$ do not change signs when the sample moves from upstream to downstream, as their features differ from that exhibited in Fig.~\ref{fig-p1=p2}(e)-(f). In fact, this difference is exactly predicted by Eq.~(\ref{eq-phi-dual-g}).  In fact, because of the different interferometer configuration, $\pfr$ and $\p0$ in these simulations are changed to  $\pfr=-191.4\mu\mathrm{m}<0$ and $\p0=11.67\mu\mathrm{m}>0$. Equation~(\ref{eq-phi-dual-g}) dictates that $\Delta\phi(x)$ will not change sign when sample is moved from upstream of $G_1$ to downstream of $G_2$.  This example demonstrates again that Eq.~(\ref{eq-phi-dual-g}) gives accurate mathematical relation between sample phase gradient and fringe phase shift, accurate not only for predicting their magnitudes but also for their signs.  Hence Eq.~(\ref{eq-phi-dual-g}) provides a useful tool in quantitative phase contrast imaging based on  dual phase grating interferometry.

\section{Discussion and conclusions}\label{sec-conclude}
One important task in x-ray interferometry is to rigorously derive the functional relationship between measured interference fringe shifts and sample phase gradients. Combined with tomography, the retrieved sample gradients from different angular projections can be used to reconstruct quantitative volumetric images of sample electron densities. The mathematical relationship between fringe shifts and sample phase gradients depends not only on an interferometer's grating periods and geometric configuration, but also on the sample position within the interferometer.   The formulas for retrieving sample phase gradient in Talbot-Lau interferometry are well known as summarized in Eq.~(\ref{eq-phi-1g}).  However, there is a need to provide corresponding formulas applicable to dual phase grating interferometry with various sample position. In this work we fulfill this need using a novel intuitive analysis based on the fringe pattern formation mechanism.  The derived formulas of Eq.~(\ref{eq-phi-dual-g}) provides a useful tool for retrieving sample phase gradients in dual phase grating interferometry with any sample position. These formulas have been validated by extensive simulations, not only in the magnitudes but also in the signs of the retrieved sample phase gradients, as is shown in section~\ref{sec-results}. In addition Tab.~\ref{tab-1} shows that sample position has a significant effect on angular sensitivity of a setup.

Moreover, the formulas of Eq.~(\ref{eq-phi-dual-g}) can also be used to compute the angular sensitivity $\Delta\phi/(2\pi\alpha)$, which represents the fringe phase shift generated per unit sample-refraction angle in a given setup. As is discussed in section~\ref{sec-results}, although it is generally desirable to have a large $\Delta\phi/(2\pi\alpha)$ for a setup, but it is only a nominal sensitivity measure. The true sensitivity of a interferometer setup is the minimum  detectable refraction angle, which depends on the fringe visibility and photon quantum noise with the setup. Detail discussion of this topic is out  the scope of this paper.

In Eq.~(\ref{eq-phi-dual-g}), the proportion constant $\xi$  between fringe phase shift and sample phase gradient is also called the auto-correlation length of an interferometer setup.  This quantity $\xi$ characterize the setup's  sensitivity for detecting small angle x-ray scattering from the sample's fine structures.  For detail readers are referred to dark field imaging literature [ref]. We found that the auto-correlation length $\xi$ is inversely proportional to fringe period $\pfr$ of Eq.~(\ref{eq-pfr}),  if the sample is positioned downstream of $G_2$ grating.  On the other hand, when the sample is positioned upstream of $G_1$, the auto-correlation length $\xi$ is inversely proportional to a length-quantity $\p0$ of Eq.~(\ref{eq-p0}). The size of $\p0$ has a physical meaning in a different context. Namely, if a source grating were incorporated, then the periodicity of the source grating should be set to $\p0$ for achieving good fringe visibility\cite{Y-W-L-dual-grating-Lau-condition}. 

In conclusion, in this work, using intuitive analysis of the sample-generated fringe shifts based on the beat pattern formation mechanism, the authors derived the formulas relating sample's phase gradients to fringe phase shifts.  These formulas provide also a design optimization tool for dual phase grating interferometry.

\section*{Funding}
National Institutes of Health (NIH) (1R01CA193378). 

\end{document}